\pdfoutput=1

\documentclass[submission,copyright]{eptcs}
\providecommand{\event}{NMCA 2023}

\usepackage{times}
\usepackage{latexsym}

\usepackage[T1]{fontenc}

\usepackage[utf8]{inputenc}
\usepackage[most]{tcolorbox}

\usepackage{tikz}
\usetikzlibrary{arrows}
\usetikzlibrary{positioning}
\usetikzlibrary{decorations.pathreplacing}
\usetikzlibrary{fit}
\usetikzlibrary{backgrounds}
\usetikzlibrary{automata}
\usetikzlibrary{positioning}
\usetikzlibrary{arrows}
\usetikzlibrary{calc}
\usetikzlibrary{decorations.markings}
\usetikzlibrary{shapes.geometric}
\usepackage{qtree}

\usepackage{amsmath,amssymb,amsfonts}
\usepackage{graphicx}
\usepackage{url}
\usepackage{mathtools}
\usepackage{bbold}
\usepackage[inline]{enumitem}
\setlist{noitemsep}
\usepackage{algorithm}
\usepackage{algpseudocode}
\usepackage{todonotes}
\usepackage{booktabs}
\usepackage{multicol}
\usepackage{xspace}
\usepackage{comment}
\usepackage{todonotes}
\usepackage{listings}
\newcommand{\lovelace}{\textsc{Lovelace}\xspace}

\title{Generating Semantic Graph Corpora with\\[.5ex] Graph Expansion Grammar 
} 

\author{Eric Andersson \qquad Johanna Björklund\thanks{Supported by the Swedish Research Council under grant number 2020-03852} \qquad Frank Drewes\thanks{Supported by the Wallenberg AI, Autonomous Systems and Software Program through the NEST project \emph{STING}} \qquad Anna Jonsson
\institute{Department of Computing Science\\Umeå University, Umeå, Sweden}
\email{\{dv20ean,\,johanna,\,drewes,\,aj\}@cs.umu.se}
}

\def\titlerunning{Generating Semantic Graph Corpora}
\def\authorrunning{Andersson et al.}

\newtheorem{theorem}{Theorem}

\newtheorem{definition}[theorem]{Definition}

\definecolor{UmUBlue}{RGB}{42,71,101}
\definecolor{UmUGreen}{RGB}{115,167,144}
\definecolor{UmUGold}{RGB}{215,177,124}
\definecolor{UmUPink}{RGB}{234,186,185}
\definecolor{NoUmUColour}{RGB}{84,142,202}

\definecolor{UmUBrightBlue}{RGB}{42,71,191}

\tikzstyle{overtex}=[ellipse,line width=0.30mm,fill=white,draw,inner sep=.5pt, minimum size=2ex]
\tikzstyle{terminal}=[line width=0.25mm,rounded corners, draw, fill=UmUGold!70]
\tikzstyle{eval}=[rounded corners, draw, line width=0mm, fill=UmUBlue!20]

\let\origQed\qed
\newcommand*{\resetqed}{\gdef\qed{\origQed\global\let\qed\relax}}
\resetqed
\let\origendproof\endproof
\def\endproof{\origendproof\resetqed}

\newcounter{cl}

\newcounter{cse}

\newcommand{\nat}{\mathbb N}
\newcommand{\pow}[1]{\wp(#1)}
\newcommand{\emptystr}{\varepsilon}

\newcommand{\norep}[1]{#1^\circledast}

\newcommand{\sig}{\Sigma}
\newcommand{\trees}[1]{\mathrm T_{#1}}

\newcommand{\alg}{\mathcal A}
\newcommand{\A}{\mathbb A}
\newcommand{\val}{\mathit{val}}
\newcommand{\rank}{\mathit{rk}}

\newcommand{\type}{\mathit{type}}

\newcommand{\emptygraph}{\phi}
\newcommand{\labels}{\mathbb L}
\newcommand{\ndlab}{\!\dot{\,\mathbb L}}
\newcommand{\edlab}{\bar{\mathbb L}}
\newcommand{\lab}{\mathit{lab}}
\newcommand{\port}{\mathit{port}}
\newcommand{\dock}{\mathit{dock}}

\newcommand{\union}[2]{\uplus_{#1#2}}

\newcommand{\context}{C}
\newcommand{\New}{\mathit{NEW}}
\newcommand{\graphs}{\mathbb G}

\newcommand{\G}{\mathcal G}
\newcommand{\und}[1]{\underline{#1}}
\newcommand{\ext}{\Phi}

\newcommand{\prt}[1]{{\mathversion{bold}\scriptsize\color{UmUBlue}$#1$}}
\newcommand{\dk}[1]{{\mathversion{bold}\scriptsize\color{UmUGreen}$(#1)$}}

\newtcolorbox{myframe}[1][]{
  enhanced,
  arc=0pt,
  outer arc=0pt,
  colback=white,
  boxrule=0.8pt,
  #1
}

\hypersetup{
    colorlinks = true,
    linkcolor = {UmUBrightBlue},
    anchorcolor = {UmUBrightBlue},
    citecolor = {UmUBrightBlue},
    linkbordercolor = {UmUBrightBlue},
    urlcolor = {UmUBrightBlue}
}

\begin{document}
\maketitle

\begin{abstract}
We introduce \lovelace{}, a tool for creating corpora of semantic graphs.
The system uses graph expansion grammar as  a representational language, thus allowing users to craft a grammar that describes a corpus with desired properties. 
When given such grammar as input, the system generates a set of output graphs that are well-formed according to the grammar, i.e., a graph bank.
The generation process can be controlled via a number of configurable parameters that allow the user to, for example, specify a range of desired output graph sizes.
Central use cases are the creation of synthetic data to augment existing corpora, and as a pedagogical tool for teaching formal language theory. 
\end{abstract}

\section{Introduction}
Semantic representations are formalisms designed to express the meaning of natural language data in a clear and concise way, which is suitable both for manual inspection and for automated processing. 
A wide range of representational formats has been considered in literature. Some of the more commonly used are based on graphs, in which nodes correspond to concepts, and edges to relations between them. Prominent examples are combinatory categorial grammar~\cite{SteedmanBaldridge:2011}, abstract meaning representation (AMR)~\cite{Langkilde:1998,BanarescuEtAl:2013} and universal conceptual cognitive annotation~\cite{AbendRappoport:2013}.

It would be valuable for many applications if one could automatically translate natural language sentences into semantic graphs. 
However, for developing, training, and testing such approaches, corpora like the AMR corpora\footnote{\url{https://amr.isi.edu}} are required. 
The creation of high-quality corpora is work intensive and requires both linguistic knowledge and a familiarity with the representational formalism at hand. 
Moreover, even skilled annotators tire, and hence the resulting translations are bound to contain errors and inconsistencies. 
In addition to this, hand-annotated real-world data is of limited use for conducting controlled experiments whose purpose it is to study the influence of particular structural properties of the representation on a given machine learning technique.

To address these problems, we provide the software \lovelace{}\footnote{\url{https://github.com/tm11ajn/lovelace/}} that generates well-formed graphs with respect to a \emph{graph expansion grammar} (GEG)~\cite{bjorklund-etal-2023}. 
GEGs are hyperedge replacement grammars~\cite{Bauderon-Courcelle:87,Habel:92,Drewes-Habel-Kreowski:97} that have been extended by a type of contextual rules inspired by~\cite{DrewesEtAl:2010}.

Technically, a GEG is defined as a regular tree grammar that generates terms over a particular graph algebra, and these terms are then evaluated into a set of directed acyclic graphs. 
As usual, the evaluation of a term is done recursively. 
Assuming that a given subterm has already been evaluated to a graph, which will become a subgraph of the generated graph, the evaluation of an operation on top of it adds new nodes with edges pointing to already existing nodes of the subgraph. 
In \cite{bjorklund-etal-2023}, the placement of these edges can be restricted by a formula in counting monadic second-order logic. 
\lovelace does not currently make use of such a powerful mechanism, which we leave for future extensions. 
In another respect (to be discussed in Section~\ref{sec:geg}), we generalise expansion operations slightly, which ensures that the formalism becomes more powerful than hyperedge replacement. 
This deviation from the original definition~\cite{bjorklund-etal-2023} is motivated by the fact that the focus in that work was on polynomial parsing, whereas \lovelace is a generative tool for which the well-known NP-completeness of hyperedge replacement languages is of no relevance. 

There are several semantically annotated treebanks available, including PropBank~\cite{palmer-etal-2005-proposition}, FrameNet~\cite{baker-etal-1998-berkeley-framenet}, and the Penn Discourse TreeBank~\cite{Prasad:2008,prasad-etal-2008-penn}. 
There are also tools that generate synthetic treebanks from grammars, which can, if so designed, contain semantic information. 
In this category of tools we have  Grammatical Framework~\cite{ranta2004grammatical}, a programming language specifically designed for writing string grammars, but which also provides functionality for generating corpora of parse trees with respect to a given grammar. 
Another example is Tiburon~\cite{May2006}, a capable toolkit for processing weighted automata which includes an algorithm for extracting $N$ parse trees with optimal weight from a weighted string grammar. 
Finally we have \textsc{Betty}~\cite{bjorklund-etal-2022-improved}, which can extract both the $N$ best derivation trees, but also the $N$ best output trees with respect to a tree grammar; cf.~Section~\ref{sec:lovelace}. 

Turning specifically to graph banks, Hockenmeir and Steedman propose an algorithm for translating the Penn Treebank into a corpus of CCG derivations augmented with local and long-range word–word dependencies~\cite{hockenmaier-steedman-2007-ccgbank}. There is also the manually created AMR bank by~\cite{Banarescu:2014}. The present paper adds to this line of work by providing a method of creating synthetic corpora of semantic graphs from a specification given in the form of a graph expansion grammar. 

The paper contains the following main sections: Section~\ref{sec:geg} recalls the graph expansion grammar formalism, Section~\ref{sec:lovelace} explains how to find and use the software, and Section~\ref{sec:conclusion} provides a summary of the work presented here together with ideas for improvement.

\section{Graph Expansion Grammar}\label{sec:geg}

To recall the graph expansion grammar formalism~\cite{bjorklund-etal-2023}, we first fix a few standard definitions and related notation from discrete mathematics and automata theory. 

The set of natural numbers (including $0$) is denoted by $\nat$, and $[n] = \{1,\dots,n\}$ for $n\in\nat$.
The set of all strings (that is, finite sequences) over a set $S$ is $S^*$, which in particular contains the empty string~$\emptystr$. The subset of $S^*$ containing only those strings which do not have repeating elements is $\norep S$.
For a string $w$, we let $[w]$ denote the smallest set $S$ such that $w\in S^*$. 
We denote the canonical extensions of a function $f\colon S\to T$ to $S^*$ and to the powerset $\pow S$ of $S$ also by $f$, i.e., $f(s_1\cdots s_n)=f(s_1)\cdots f(s_n)$ for $s_1,\dots,s_n\in S$, and $f(S')=\{f(s)\mid s\in S'\}$ for $S'\in \pow S$. 

A \emph{ranked alphabet} is a pair $A=(\sig,\rank)$ consisting of a finite set of symbols $\sig$ and a function $\rank\colon\sig\to\nat$ that assigns a rank to every symbol $\sigma\in\sig$. 
Writing $\sigma^{(k)}$ indicates that $\rank(\sigma)=k$. If there is no danger of confusion, we keep $\rank$ implicit and identify $A$ with $\sig$.

The set $\trees\sig$ of all \emph{trees over $\sig$} is the smallest set of formal expressions such that $\sigma[t_1,\dots,t_k]\in\trees\sig$ for every $\sigma^{(k)}\in\sig$ and all trees $t_1,\dots,t_k\in\trees\sig$.  
Thus, the rank~$k$ of $\sigma$ determines the number of subtrees of every occurrence of~$\sigma$ in a tree. If $k=0$, then $f[]\in\trees\sig$, which we abbreviate as $f$, omitting the brackets.

Given a ranked alphabet $\sig$ as above, a \emph{$\sig$-algebra} is a pair $\alg=(\A,(f_\alg)_{f\in \sig})$ consisting of a set $\A$, the \emph{domain} of~$\alg$, and a function $f_\alg\colon \A^k\to\A$ for every $f^{(k)}\in\sig$, the \emph{interpretation} of $f$ in $\alg$. 
Now, if $t=f[t_1,\dots,t_k]$ is a tree in $\trees\sig$, evaluating $t$ with respect to $\alg$ yields $\val_\alg(t)\in\A$, defined as $\val_\alg(t)=f_\alg(\val_\alg(t_1),\dots,\val_\alg(t_k))$.

To generate trees over the operations of an algebra, we use \emph{regular tree grammars}. 

\begin{definition}\label{de:rtg}
A \emph{regular tree grammar (over $\sig$)} is a tuple $g=(N,\sig,P,S)$ consisting of
\begin{itemize}
    \item a ranked alphabet $N$ of symbols of rank~$0$, called \emph{nonterminals},
    \item a ranked alphabet $\sig$ of \emph{terminals}, disjoint with $N$,
    \item a set $P$ of \emph{productions} $A\to f[A_1,\dots,A_k]$ where $f^{(k)}\in\sig$ for some $k\in\nat$ and $A,A_1,\dots,A_k\in N$, and
    \item an \emph{initial nonterminal} $S\in N$.
\end{itemize}

The \emph{regular tree language} (rtg) generated by $g$ is $L(g)=L_S(g)$ where $(L_A(g))_{A\in N}$ is the smallest family of subsets of \hspace{2pt}$\trees\sig$ such that, for $A\in N$, a tree $f[t_1,\dots,t_k]$ is in $L_A(g)$ if $(A \to f[A_1,\dots,A_k])\in P$ and $t_i\in L_{A_i}(g)$ for all $i\in[k]$. (See Figures~\ref{fig:rtg} and \ref{fig:tree} for an example regular tree grammar and a tree in its language, respectively.)
\end{definition}

To generate languages other than tree languages using regular tree grammars, we follow the idea of the seminal paper by Mezei and Wright~\cite{Mezei-Wright:67}: the combination of a regular tree grammar $g$ over $\sig$ and a $\sig$-algebra $\alg$ generates the subset of~$\A$ whose elements are all $\val_\alg(t)$ such that $t\in L(g)$. In our case, $\A$ is the set of graphs (over a given set of labels). The operations are, thus, operations on graphs. However, the central operation is nondeterministic, meaning that its application to a given graph can produce several possible outputs. Formally, we model this by letting the operations work on sets of graphs instead of individual graphs.

The graphs we work with are node- and edge-labelled directed graphs, each equipped with a sequence of so-called ports. 
From a graph operation point of view, the sequence of ports of a graph is its ``interface'': its nodes are the only ones that can individually be accessed by operations to attach new edges to them. The number of ports is the \emph{type} of the graph.

\begin{definition}\label{def:graph}
Let $\labels=(\ndlab,\edlab)$ be a \emph{labelling alphabet}: a pair of finite sets of labels $\ndlab$ and $\edlab$. A \emph{graph} over $\labels$ is a tuple $G=(V,E,\lab,\port)$ such that
\begin{itemize}
    \item $V$ is the finite set of \emph{nodes},
    \item $E\subseteq V\times\edlab\times V$ is the set of \emph{edges},
    \item $\lab\colon V\to\ndlab$ labels the nodes, and
    \item $\port\in\norep V$ is the sequence of \emph{ports} of the graph.
\end{itemize}
The \emph{type} of $G$ is $\type(G)=|\port|$. 
The set of all graphs of type $k$ is denoted by $\graphs_k$. 
\end{definition}

If the components of a graph $G$ are not explicitly named, they are denoted by $V_G$, $E_G$, $\lab_G$, and $\port_G$, respectively.


Graph expansion grammars generate graphs using two types of graph operations: disjoint union and the more complex graph expansion operations.
Disjoint union just combines two graphs into one by placing them next to each other (after making their node sets disjoint) and concatenating their port sequences. 
Formally, let $k,k'\in\nat$. 
Then $\union k{k'}\colon\graphs_k\times\graphs_{k'}\to\graphs_{k+k'}$ is defined as follows: for $G\in\graphs_k$ and $G'\in\graphs_{k'}$ with disjoint sets of nodes, $\union k{k'}(G,G')$ yields the graph $(V,E,\lab,\port)\in\graphs_{k+k'}$ given by $V=V_G\cup V_{G'}$, $E=E_G\cup E_{G'}$, $\lab=\lab_G\cup\lab_{G'}$, and $\port=\port_G\port_{G'}$.\footnote{Here, $\lab_G\cup\lab_{G'}$ is the usual union of binary relations.} If $V_G\cap V_{G'}\neq\emptyset$, we silently rename nodes before we apply $\union k{k'}$, because we are only interested in generating graphs up to isomorphism.
Note that $\union k{k'}$ is not commutative because of the concatenation of port sequences. 
We usually write $G\union k{k'}G'$ instead of $\union k{k'}(G,G')$.
We extend $\union k{k'}$ to $\union k{k'}\colon\pow{\graphs_k}\times\pow{\graphs_{k'}}\to\pow{\graphs_{k+k'}}$ by letting $\G\union k{k'}\G'=\{G\union k{k'}G'\mid G\in\G,\ G'\in\G'\}$ for $\G\subseteq\graphs_k$ and $\G'\subseteq\graphs_{k'}$.

The other type of operation, the \emph{graph expansion}, extends an existing graph with an additional structure placed ``on top'' of that graph. 
Expansion is specified by a template graph with an additional sequence of designated nodes called \emph{docks}. 
Applying an extension operation adds the template graph to the argument graph and identifies the docks with the ports of that graph. The ports of the template become the ports of the combined graph. 
The template also contains a number of \emph{context nodes} that can be identified with arbitrarily chosen nodes with matching labels in the argument graph. 
Formally, a graph expansion operation is a unary operation given by a tuple
$\ext=(V,E,\lab,\port,\dock)$
where $(V,E,\lab,\port)$, henceforth denoted by $\und\ext$, is the \emph{underlying graph} and $\dock\in V^*$ is the sequence of \emph{docks}. Note that $\dock$, in contrast to $\port$, may contain repetitions. Similarly to our notation for the components of graphs, we use the notations $V_\ext$, $E_\ext$, $\lab_\ext$, $\port_\ext$, and $\dock_\ext$ if these components are not explicitly named. Furthermore, we let $\context_\ext=V\setminus([\port]\cup[\dock])$ denote the set of \emph{context nodes} of $\ext$.

An expansion operation $\ext$ as above can be applied to an argument graph $G=(V,E,\lab,\port)\in\graphs_\ell$ if $|\dock_\ext|=\ell$. It then yields a graph of type $|\port_\ext|$ by identifying the nodes in $\dock_\ext$ with those in $\port$, and each context node with an arbitrary node in $V$ that carries the same label. 
The port sequence of the resulting graph is $\port_\ext$.

Formally, let $|\port_\ext|=k$ and $|\dock_\ext|=\ell$. 
Then $\ext$ is interpreted as the nondeterministic operation $\ext\colon\graphs_\ell\to\pow{\graphs_k}$ defined as follows. 
For a graph $G=(V,E,\lab,\port)\in\graphs_\ell$, a graph $H\in \graphs_k$ is in $\ext(G)$ if it can be obtained by the following stepewise procedure:
\begin{enumerate}
    \item Rename the nodes of $\ext$ to make the set of nodes of $\ext$ disjoint with $V$. (As in the case of $\union{}{}$, we will in the following assume that this is done silently ``under the hood''.)
    \item Add the nodes and edges of $\und\ext$ to $G$.
    \item Identify the $i$-th node $v$ of $\port$ with the $i$-th node of $\dock_\ext$ for all $i\in[\ell]$ and label the resulting node with $\lab_\ext(v)$.
    \item Identify every node $u\in\context_\ext$ with any node $v\in V\setminus[\port]$ for which $\lab(v)=\lab_\ext(u)$.
    \item Define $\port_H=\port_\ext$.
\end{enumerate}

Note that the process of identifying docks of $\ext$ with ports of the argument graph~$G$ may merge ports of $G$ if $\dock$ contains repetitions. The expansion operations defined here are thus more general than those in \cite{bjorklund-etal-2023}. In fact, readers familiar with hyperedge replacement grammars will easily be able to see that this allows us to simulate hyperedge replacement. Together with the fact that context nodes can be used to create graphs of unbounded treewidth, this implies that graph expansion grammars, to be defined below, are strictly more powerful than hyperedge replacement grammars.

Further deviations from~\cite{bjorklund-etal-2023} are that the definition above does not make use of the cloning of context nodes, and that the logic formula that determines which mappings of context nodes to nodes in the argument graph are allowed has been replaced by the much simpler condition that node labels must match. The cloning ability is not needed here since we consider expansion operations rather than the special case of extension operations as in~\cite{bjorklund-etal-2023} (see below), which means that cloning can be implemented by repeated application of expansion. The latter has been dropped in the current paper for simplicity, and because it is not yet implemented in \lovelace anyway.

The major result of \cite{bjorklund-etal-2023} applies to a restricted form of expansion operations, the so-called extension operations. By using only extension operations, we can make sure that graphs are built bottom-up, that is, that $\ext$ always extends the input graph by placing nodes and edges ``on top'', with edges being directed downwards, and that all nodes of generated graphs are reachable from the ports. For a brief explanation, let $\New_\ext=[\port_\ext]\setminus[\dock_\ext]$ denote the set of nodes that an application of $\ext$ adds to the graph, i.e.~those nodes of $\und\ext$ which are not identified with nodes of the argument graph when $\ext$ is applied. Then $\ext$ is an extension operation if it satisfies the following requirements:
\begin{enumerate}[label=(R\arabic*),leftmargin=*,series=R]
    \item\label{sources} $E_\ext\subseteq\New_\ext\times\edlab\times(V_\ext\setminus\New_\ext)$ and
    \item\label{forget} every node in $[\dock_\ext]\setminus[\port_\ext]$ has an incoming edge.
\end{enumerate}

By induction, \ref{sources} ensures that all graphs generated by a graph extension grammar (i.e., a GEG all of whose expansion operations are extension operations) are directed acyclic graphs.
Likewise by induction, \ref{forget} ensures that every node in a graph generated by a graph extension grammar is reachable from a port. While these restrictions are not employed in the current paper (since they are not needed unless one is interested in efficient parsing), they are well justified when generating semantic graphs such as AMR, because these typically consist of directed acyclic graphs in which all nodes are reachable from the roots (which would translate to ports in the graph grammar formalism). Thus, while \lovelace does not enforce~\ref{sources} and~\ref{forget}, our examples will actually obey these requirements.

\begin{figure*}[htb]
    \centering
    \tikzset{
  every label/.style={draw=none, fill=none, inner sep=1pt}
}
     \begin{tikzpicture}[xscale=.5, yscale=.75]
     \tikzstyle{overtex}=[circle,line width=0.30mm,fill=white,draw,inner sep=.5pt, minimum size=2ex];
          
     \input{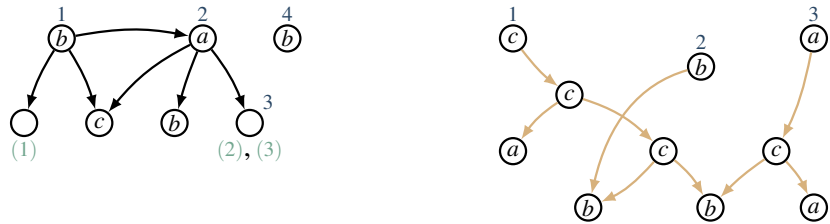}
     
     \begin{scope}[shift={(13,4)}]
     \node at (0,-2.5) [overtex, label=above:\prt 1] (Xz1) {\footnotesize $c$};
     \node at (8,-2.5) [overtex, label=above:\prt 3] (Xz2) {\footnotesize $a$};
     
     
     \node at (1.5,-3.5) [overtex] (Xu1) {\footnotesize $c$};
     \node at (5,-3) [overtex, label=above:\prt 2] (Xu2) {\footnotesize $b$};
     
     \node at (0,-4.5) [overtex] (Xv1) {\footnotesize $a$};
     \node at (4,-4.5) [overtex] (Xv2) {\footnotesize $c$};
     \node at (7,-4.5) [overtex] (Xv3) {\footnotesize $c$};
     
     \node at (2,-5.5) [overtex] (Xw1) {\footnotesize $b$};
     \node at (5.25,-5.5) [overtex] (Xw2) {\footnotesize $b$};
     \node at (8,-5.5) [overtex] (Xw3) {\footnotesize $a$};
     
    \draw[->,line width=0.30mm,-latex,UmUGold,bend right=7] (Xz1) to (Xu1);
    \draw[->,line width=0.30mm,-latex,UmUGold,bend left=10] (Xz2) to (Xv3);
    \draw[->,line width=0.30mm,-latex,UmUGold,bend right=10] (Xu1) to (Xv1);
    \draw[->,line width=0.30mm,-latex,UmUGold,bend left=10] (Xu1) to (Xv2);
    \draw[->,line width=0.30mm,-latex,UmUGold,bend right=30] (Xu2) to (Xw1);

    \draw[->,line width=0.30mm,-latex,UmUGold,bend left=10] (Xv2) to (Xw1);
    \draw[->,line width=0.30mm,-latex,UmUGold,bend left=10] (Xv2) to (Xw2);
    \draw[->,line width=0.30mm,-latex,UmUGold,bend right=5] (Xv3) to (Xw2);
    \draw[->,line width=0.30mm,-latex,UmUGold,bend left=10] (Xv3) to (Xw3);
    
    \path [use as bounding box] ;
    
    \end{scope}
    \end{tikzpicture}
    
    \caption{The figure on the left shows an expansion operation $\ext$ with four ports (indicated with numbers above the nodes), three docks (indicated with numbers in parentheses below the nodes), and two context nodes (the ones that are neither ports nor docks). Docks~2 and~3 coincide. Applying the expansion operation identifies docks with corresponding ports of the argument graph and each context node with a non-port in the input graph that carries a matching label. The application of $\ext$ to the graph $G$ (on the right) yields a non-empty number of possible results because the number of ports of $G$ coincides with the number of docks of $\ext$, and since there are nodes labelled $b$ and $c$ in $G$ which are not ports.}
    \label{fig:expansion_operation_and_graph}
\end{figure*}

\begin{figure*}[htb]
    \centering
    \tikzset{
  every label/.style={draw=none, fill=none, inner sep=1pt}
}
     \begin{tikzpicture}[xscale=.45, yscale=.75]
     \tikzstyle{overtex}=[circle,line width=0.30mm,fill=white,draw,inner sep=.5pt, minimum size=2ex];

     \node at (1,0) [overtex, label=above:\prt 1] (x1) {\footnotesize $b$};
     \node at (4.75,0) [overtex, label=above:\prt 2] (x2) {\footnotesize $a$};
     \node at (7,0) [overtex, label=above:\prt 4] (x3) {\footnotesize $b$};
     

     \node at (0,-2.5) [overtex] (Xz1) {\footnotesize $c$};
     
     
     \node at (1.5,-3.5) [overtex] (Xu1) {\footnotesize $c$};
     \node at (5,-3) [overtex, label=above left:\prt 3] (Xu2) {\footnotesize $b$};
     
     \node at (0,-4.5) [overtex] (Xv1) {\footnotesize $a$};
     \node at (4,-4.5) [overtex] (Xv2) {\footnotesize $c$};
     \node at (7,-4.5) [overtex] (Xv3) {\footnotesize $c$};
     
     \node at (2,-5.5) [overtex] (Xw1) {\footnotesize $b$};
     \node at (5.25,-5.5) [overtex] (Xw2) {\footnotesize $b$};
     \node at (8,-5.5) [overtex] (Xw3) {\footnotesize $a$};
     
    \draw[->,line width=0.30mm,-latex,bend left=7] (x1) to (x2);
    \draw[->,line width=0.30mm,-latex,bend right=10] (x1) to (Xz1);
    \draw[->,line width=0.30mm,-latex] (x2) .. controls +(3.5,-2.5) and +(.5,3) .. (Xw2);
    \draw[->,line width=0.30mm,-latex,bend left=20] (x2) to (Xu2);
     
    \draw[->,line width=0.30mm,-latex,UmUGold,bend right=7] (Xz1) to (Xu1);
    \draw[->,line width=0.30mm,-latex,UmUGold,bend left=10] (Xu2) to (Xv3);
    \draw[->,line width=0.30mm,-latex,UmUGold,bend right=10] (Xu1) to (Xv1);
    \draw[->,line width=0.30mm,-latex,UmUGold,bend left=10] (Xu1) to (Xv2);
    \draw[->,line width=0.30mm,-latex,UmUGold,bend right=30] (Xu2) to (Xw1);
    \draw[->,line width=0.30mm,-latex,UmUGold,bend left=10] (Xu2) to (Xv3);
    \draw[->,line width=0.30mm,-latex,bend left=10] (x1) to (Xu1);
    \draw[->,line width=0.30mm,-latex,bend right=10] (x2) to (Xu1);

    \draw[->,line width=0.30mm,-latex,UmUGold,bend left=10] (Xv2) to (Xw1);
    \draw[->,line width=0.30mm,-latex,UmUGold,bend left=10] (Xv2) to (Xw2);
    \draw[->,line width=0.30mm,-latex,UmUGold,bend right=5] (Xv3) to (Xw2);
    \draw[->,line width=0.30mm,-latex,UmUGold,bend left=10] (Xv3) to (Xw3);

    \path [use as bounding box] ;
    
    \end{tikzpicture}
    \qquad
    \tikzset{
  every label/.style={draw=none, fill=none, inner sep=1pt}
}
     \begin{tikzpicture}[xscale=.45, yscale=.75]
     \tikzstyle{overtex}=[circle,line width=0.30mm,fill=white,draw,inner sep=.5pt, minimum size=2ex];

     \node at (1,0) [overtex, label=above:\prt 1] (x1) {\footnotesize $b$};
     \node at (4.75,0) [overtex, label=above:\prt 2] (x2) {\footnotesize $a$};
     \node at (7,0) [overtex, label=above:\prt 4] (x3) {\footnotesize $b$};
     

     \node at (0,-2.5) [overtex] (Xz1) {\footnotesize $c$};
     
     
     \node at (1.5,-3.5) [overtex] (Xu1) {\footnotesize $c$};
     \node at (5,-3) [overtex, label=above left:\prt 3] (Xu2) {\footnotesize $b$};
     
     \node at (0,-4.5) [overtex] (Xv1) {\footnotesize $a$};
     \node at (4,-4.5) [overtex] (Xv2) {\footnotesize $c$};
     \node at (7,-4.5) [overtex] (Xv3) {\footnotesize $c$};
     
     \node at (2,-5.5) [overtex] (Xw1) {\footnotesize $b$};
     \node at (5.25,-5.5) [overtex] (Xw2) {\footnotesize $b$};
     \node at (8,-5.5) [overtex] (Xw3) {\footnotesize $a$};
     
    \draw[->,line width=0.30mm,-latex,bend left=7] (x1) to (x2);
    \draw[->,line width=0.30mm,-latex,bend right=10] (x1) to (Xz1);
    \draw[->,line width=0.30mm,-latex,bend left=10] (x1) to (Xv2);
    \draw[->,line width=0.30mm,-latex] (x2) .. controls +(3.5,-2.5) and +(.5,3) .. (Xw2);
    \draw[->,line width=0.30mm,-latex,bend right=10] (x2) to (Xv2);
    \draw[->,line width=0.30mm,-latex,bend left=20] (x2) to (Xu2);
     
    \draw[->,line width=0.30mm,-latex,UmUGold,bend right=7] (Xz1) to (Xu1);
    \draw[->,line width=0.30mm,-latex,UmUGold,bend left=10] (Xu2) to (Xv3);
    \draw[->,line width=0.30mm,-latex,UmUGold,bend right=10] (Xu1) to (Xv1);
    \draw[->,line width=0.30mm,-latex,UmUGold,bend left=10] (Xu1) to (Xv2);
    \draw[->,line width=0.30mm,-latex,UmUGold,bend right=30] (Xu2) to (Xw1);
    \draw[->,line width=0.30mm,-latex,UmUGold,bend left=10] (Xu2) to (Xv3);

    \draw[->,line width=0.30mm,-latex,UmUGold,bend left=10] (Xv2) to (Xw1);
    \draw[->,line width=0.30mm,-latex,UmUGold,bend left=10] (Xv2) to (Xw2);
    \draw[->,line width=0.30mm,-latex,UmUGold,bend right=5] (Xv3) to (Xw2);
    \draw[->,line width=0.30mm,-latex,UmUGold,bend left=10] (Xv3) to (Xw3);
    
    \path [use as bounding box] ;
    
    \end{tikzpicture}
    \qquad
    \tikzset{
  every label/.style={draw=none, fill=none, inner sep=1pt}
}
     \begin{tikzpicture}[xscale=.45, yscale=.75]
     \tikzstyle{overtex}=[circle,line width=0.30mm,fill=white,draw,inner sep=.5pt, minimum size=2ex];

     \node at (1,0) [overtex, label=above:\prt 1] (x1) {\footnotesize $b$};
     \node at (4.75,0) [overtex, label=above:\prt 2] (x2) {\footnotesize $a$};
     \node at (7,0) [overtex, label=above:\prt 4] (x3) {\footnotesize $b$};
     

     \node at (0,-2.5) [overtex] (Xz1) {\footnotesize $c$};
     
     
     \node at (1.5,-3.5) [overtex] (Xu1) {\footnotesize $c$};
     \node at (5,-3) [overtex, label=above left:\prt 3] (Xu2) {\footnotesize $b$};
     
     \node at (0,-4.5) [overtex] (Xv1) {\footnotesize $a$};
     \node at (4,-4.5) [overtex] (Xv2) {\footnotesize $c$};
     \node at (7,-4.5) [overtex] (Xv3) {\footnotesize $c$};
     
     \node at (2,-5.5) [overtex] (Xw1) {\footnotesize $b$};
     \node at (5.25,-5.5) [overtex] (Xw2) {\footnotesize $b$};
     \node at (8,-5.5) [overtex] (Xw3) {\footnotesize $a$};
     
    \draw[->,line width=0.30mm,-latex,bend left=7] (x1) to (x2);
    \draw[->,line width=0.30mm,-latex,bend right=10] (x1) to (Xz1);

    \draw[->,line width=0.30mm,-latex,bend left=30] (x1) .. controls +(3,-1) and +(-.5,3) .. (Xv3);

    \draw[->,line width=0.30mm,-latex,bend right=15] (x2) to (Xw1);
    \draw[->,line width=0.30mm,-latex,bend left=20] (x2) to (Xu2);
    \draw[->,line width=0.30mm,-latex] (x2) .. controls +(3,-1) and +(2,3) .. (Xv3);
     
    \draw[->,line width=0.30mm,-latex,UmUGold,bend right=7] (Xz1) to (Xu1);
    \draw[->,line width=0.30mm,-latex,UmUGold,bend left=10] (Xu2) to (Xv3);
    \draw[->,line width=0.30mm,-latex,UmUGold,bend right=10] (Xu1) to (Xv1);
    \draw[->,line width=0.30mm,-latex,UmUGold,bend left=10] (Xu1) to (Xv2);
    \draw[->,line width=0.30mm,-latex,UmUGold,bend right=15] (Xu2) to (Xw1);
    \draw[->,line width=0.30mm,-latex,UmUGold,bend left=10] (Xu2) to (Xv3);

    \draw[->,line width=0.30mm,-latex,UmUGold,bend left=10] (Xv2) to (Xw1);
    \draw[->,line width=0.30mm,-latex,UmUGold,bend left=10] (Xv2) to (Xw2);
    \draw[->,line width=0.30mm,-latex,UmUGold,bend right=5] (Xv3) to (Xw2);
    \draw[->,line width=0.30mm,-latex,UmUGold,bend left=10] (Xv3) to (Xw3);
    
    \path [use as bounding box] ;
    
    \end{tikzpicture}
    \caption{Three graphs in $\ext(G)$ where $\ext$ and $G$ are as in Figure~\ref{fig:expansion_operation_and_graph}. The differences between the graphs reflect how the context nodes in $\ext$ were chosen to be mapped to nodes in $G$.}
    \label{fig:resulting_graphs}
\end{figure*}
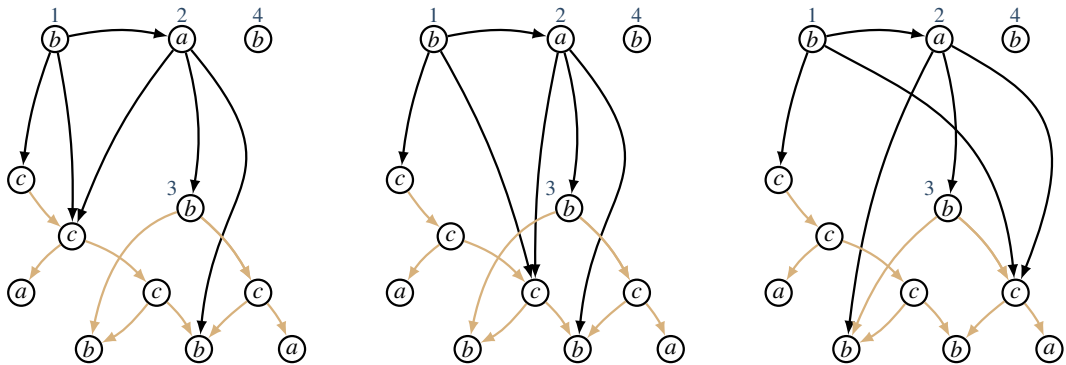

Figure~\ref{fig:expansion_operation_and_graph} depicts an expansion operation together with a graph to which it can be applied. 
Figure~\ref{fig:resulting_graphs} shows three different graphs, all resulting from the application of the expansion operation to the (now argument) graph in Figure~\ref{fig:resulting_graphs}. The resulting graphs differ because different mappings of context nodes to nodes in the argument graph were chosen. Note that $\ext$ fuses ports~$2$ and~$3$ of the argument graph, which become port~$3$ of the result, because docks~$2$ and~$3$ coincide.

A \emph{graph expansion algebra} is a $\sig$-algebra $\alg=(\pow{\graphs},(f_\alg)_{f\in\sig})$ where every symbol in $\sig$ is interpreted as an expansion operation, a union operation, or the set $\{\emptygraph\}$, where $\emptygraph$ is the empty graph $(\emptyset,\emptyset,\emptyset,\emptystr)$. 
As previously mentioned, the operations of the algebra act on sets of graphs rather than on single graphs, due to the nondeterministic nature of expansion. 
This also takes care of the fact that operations are only defined on graphs of matching types: we simply use the convention that the application of an operation to a graph of an inappropriate type returns the empty set.

\begin{definition}
A \emph{graph expansion grammar} is a pair $\Gamma=(g,\alg)$ where $\alg$ is a graph expansion $\sig$-algebra for some ranked alphabet $\sig$ and $g$ is a regular tree grammar over $\sig$.
\[
L(\Gamma)=\bigcup_{t\in L(g)}\val_\alg(t)
\] 
is the \emph{graph language generated by $\Gamma$}.
\end{definition}

\section{\lovelace{}}\label{sec:lovelace}

Let us now make use of the capacity of graph expansion grammars for expressing semantic graph languages to create a semantic graph generator.
We named the software tool that implements this functionality \lovelace{}\footnote{\url{https://github.com/tm11ajn/lovelace/}}.
To use \lovelace{}, one needs to have access to, or themselves define, a graph expansion grammar describing a language that contains the wanted corpora. 
In the rest of this section, we explain in greater detail how to combine \lovelace{} with the tool \textsc{Betty}\footnote{\url{https://github.com/tm11ajn/betty/}} to generate graph corpora.

\textsc{Betty} operates on weighted regular tree grammars, that is, on rtgs in which the rules are equipped with weights. 
In the case of \textsc{Betty}, these must be taken from the tropical semiring. 
The resulting grammars work precisely like those in Definition~\ref{de:rtg}, but assign an additional weight to every generated tree, computed as follows. 
The weight of a derivation is the sum of all weights of the rules applied to generate the tree. 
The weight of a tree in the language is the minimum of all weights of derivations that yield that tree. 
\textsc{Betty} takes as input such a weighted grammar and some natural number~$N$, and outputs~$N$ \emph{best trees}, that is, $N$~pairwise distinct trees of least weight (in the order of increasing weight). 
Thus, in this context, lesser weight is better. In the case of ties, \textsc{Betty} gives precedence to smaller trees. 
In particular, assigning all rules the same weight results in picking~$N$ smallest possible trees from the generated language. 
It is in fact unnecessary to provide rules with an explicit weight as \textsc{Betty} interprets rules without a weight as rules of weight~$0$. For simplicity, the example we use below in order to illustrate the generation of corpora makes use of this possibility.

\begin{figure}[h]
\begin{minipage}[b]{0.4\textwidth}
\begin{myframe}[width=.9\textwidth,top=5pt,bottom=5pt,left=10pt,right=10pt]

\begin{lstlisting}[xleftmargin=2em,numbers=left,stepnumber=1,breaklines=true]
S
S -> op1(C) 
C -> op2(U)
U -> op3(S' S)
S' -> op4
S -> op5
\end{lstlisting}
\end{myframe}

\caption{A regular tree grammar (on \texttt{rtg} format). The first nonterminal in the file represents the starting nonterminal.}
\label{fig:rtg}
\end{minipage}\hfill
\begin{minipage}[b]{0.55\textwidth}
\centering
\begin{tikzpicture}
    \node at (0,0) (n1) {$\mathit{op}_1$};
    \node[below of=n1] (n2) {$\mathit{op}_2$};
    \node[below of=n2] (n3) {$\mathit{op}_3$};
    \node[below left of=n3] (n4) {$\mathit{op}_4$};
    \node[below right of=n3] (n5) {$\mathit{op}_5$};
    \draw (n1) to (n2);
    \draw (n2) to (n3);
    \draw (n3) to[bend right] (n4);
    \draw (n3) to[bend left] (n5);
\end{tikzpicture}
\caption{A visual representation of the unique tree $\texttt{op1}[\texttt{op2}[\texttt{op3}[\texttt{op4}, \,\texttt{op5}]]$ in the language generated by the regular tree grammar in Figure~\ref{fig:rtg} on the left.}
\label{fig:tree}
\end{minipage}
\end{figure}

To generate the semantic graph corpora, a two-step approach is used: First $N$ best trees are extracted from the (now weighted) regular tree grammar component of the graph expansion grammar, and these are then evaluated with respect to the algebra.
As there is currently no direct integration of \textsc{Betty} and \lovelace{}, this pipeline must be set up manually. Syntactically, the input format to \textsc{Betty} is the \texttt{rtg} format of~\cite{May2006}; see that paper for more information.
An example regular tree grammar on \texttt{rtg} format can be seen in Figure~\ref{fig:rtg}, and Figure~\ref{fig:tree} shows an example tree in the corresponding language.
The trees that \textsc{Betty} then outputs are the derivation trees that comprise the basis of the corpus. 

\begin{figure}[t!]
\centering
%
%
%
%
%
\tikzset{
  every label/.style={draw=none, fill=none, inner sep=1pt}
}

\newcommand{\rhs}[2]{\left.\left.\begin{array}{@{}l@{}}#1\end{array}\right[#2\right]}

\[\begin{array}{@{}l@{\ }c@{\hspace{-1ex}}c@{\hspace{1.25cm}}l@{\ }c@{\hspace{-1ex}}c@{}}
      & & 
    \begin{tikzpicture}[xscale=.685, yscale=1]
    
    \node at (-1.25, 0.75) (name) {$\mathit{op}_1 \colon$};
    \node at (2,1.25) [terminal, inner sep=4pt, label=above:\prt 1] (x1) {\scriptsize \texttt{persuade}};
     
    \node at (0,-.35) [terminal] (y1) {\scriptsize \textrm{she}};
    \node at (2,-.35) [overtex, label=below:\dk 1] (y2) {};
    \node at (4,-.35) [overtex, label=below:\dk 2] (y3) {};
     
    \draw[->,line width=0.30mm,-latex,bend right=20] (x1) to node [pos=0.5, above, sloped] (a0) {\scriptsize \texttt{arg0}} (y1);
    \draw[->,line width=0.30mm,-latex] (x1) to node [pos=0.55, above, sloped] (a1) {\scriptsize \texttt{arg1}} (y2);
    \draw[->,line width=0.30mm,-latex,bend left=20] (x1) to node [pos=0.55, above, sloped] (a1) {\scriptsize \texttt{arg2}} (y3);
    \path [use as bounding box] ;
    \end{tikzpicture}
    &
     & & 
    \begin{tikzpicture}[xscale=.685, yscale=.90]
    \node at (-1.75, 0.75) (name) {$\mathit{op}_2 \colon$};
    \node at (1,1.25) [terminal,  label=above:\prt 2] (x1) {\scriptsize \texttt{belive}};
     
    \node at (-.5,-.35) [overtex, label=100:\prt 1, label=below:\dk 1] (y1) {};
    \node at (2.5,-.35) [overtex, label=below:\dk 2] (y2) {};
     
    \draw[->,line width=0.30mm,-latex,bend right=20] (x1) to node [pos=0.45, above, sloped] (a0) {\scriptsize \texttt{arg0}} (y1);
    \draw[->,line width=0.30mm,-latex,bend left=20] (x1) to node [pos=0.5, above, sloped] (a1) {\scriptsize \texttt{arg1}} (y2);
    \path [use as bounding box] ;
    \end{tikzpicture}%
    
\\\\[-8pt]
\multicolumn{6}{l}{
	\mathit{op}_3 \colon  S' \union11 S \qquad\qquad\qquad 
    \mathit{op}_4 \colon 
    \begin{tikzpicture}[scale=.7]
    \node at (0,0) [terminal, label=above:\prt 1] (y1) {\scriptsize they};
    \path [use as bounding box] ;
    \end{tikzpicture} \qquad\qquad\qquad 
    \mathit{op}_5 \colon 
    \begin{tikzpicture}[scale=.7]
    \node at (0,0) [terminal, label=above:\prt 1] (y1) {\scriptsize she};
    \path [use as bounding box] ;
    \end{tikzpicture}
}

\end{array}\]
\vspace{-3ex}
\caption{A definition of five graph operations. Here, $\mathit{op}_3$ is a union operation that takes two argument graphs with one port each, and the remaining operations are graph expansion operations.}
\label{fig:ops}
\end{figure}
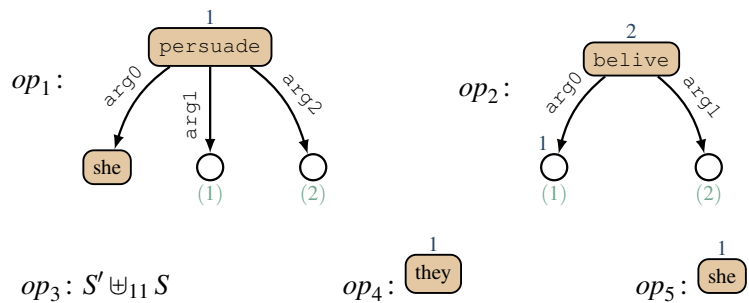

\begin{figure}
\begin{minipage}[b]{0.45\textwidth}
\begin{lstlisting}[xleftmargin=2em,numbers=left,stepnumber=1,breaklines=true]
operation op3 {
  1 1
}
\end{lstlisting}
    \caption{A textual representation of the union operation \texttt{op3} that takes two graphs with one port each and turns them into a single graph with two ports.}
    \label{fig:union-op-file-format}
\end{minipage}\hfill
\begin{minipage}[b]{0.5\textwidth}
\begin{lstlisting}[xleftmargin=2em,numbers=left,stepnumber=1,breaklines=true]
operation op1 {
  0 [label="persuade", port=1]
  1 [label="she"]
  2 [dock=1]
  3 [dock=2]
  0 -> 1 [label="arg0"]
  0 -> 2 [label="arg1"]
  0 -> 3 [label="arg2"]
}
\end{lstlisting}
    \caption{A textual representation of the expansion operation $op_1$ in Figure~\ref{fig:ops}.
    The name of the operation is \texttt{op1}, which is also the label that is used in a tree grammar file to refer to this operation.}
    \label{fig:ext-op-file-format}
\end{minipage}
\end{figure}

\begin{figure}[ht]
\begin{minipage}[b]{0.5\textwidth}
    \centering
        \begin{myframe}[width=.9\textwidth,top=5pt,bottom=5pt,left=10pt,right=10pt]
\parbox[c]{\textwidth}{
\begin{lstlisting}[xleftmargin=2em,numbers=left,stepnumber=1,breaklines=true]
digraph G  {
  0 [label="they"]
  1 [label="she"]
  2 [label="believe"]
  3 [label="persuade"]

  2 -> 0 [label="arg0"]
  2 -> 1 [label="arg1"]
  3 -> 1 [label="arg0"]
  3 -> 0 [label="arg1"]
  3 -> 2 [label="arg2"]
}
\end{lstlisting}
     }
\end{myframe}
    \caption{A file in the Graphviz format \texttt{gv} representing the graph resulting from evaluating the tree in Figure~\ref{fig:tree} with respect to the operations in Figure~\ref{fig:ops}.}
    \label{fig:text_of_dag}
\end{minipage}\hfill
\begin{minipage}[b]{0.45\textwidth}
    \centering
    \includegraphics[scale = 0.55]{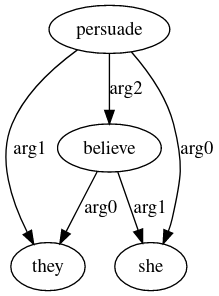}
    \caption{A visualisation of the output file depicted in Figure~\ref{fig:text_of_dag}, created using Graphviz.}
    \label{fig:DAG_viz}
\end{minipage}
\end{figure}

In the next step of the generation process, the derivation trees are translated into graphs using \lovelace{}.
To do this, we must specify the graph expansion algebra. In other words, we must associate an operation with every terminal in the regular tree grammar and gather them in an operation file. 
Such a set of operations for the tree of Figure~\ref{fig:tree} is depicted in Figure~\ref{fig:ops}.
Union operations and expansion operations have similar textual formats.
Both use the keyword \texttt{operation} together with the name of the operation (i.e., the corresponding terminal in the regular tree grammar) and curly brackets to enclose the operation specification.
A union operation is specified -- as seen in Figure~\ref{fig:union-op-file-format} -- using a single line of two numbers referring to the number of ports of the two input arguments. 
An expansion operation must necessarily specify a graph with ports and docks, which is why we found it convenient to base the representation on the \texttt{gv} digraph format used by the open-source tool Graphviz\footnote[1]{\url{https://graphviz.org/}} (see Figures~\ref{fig:text_of_dag} and \ref{fig:DAG_viz} for an example).
In Figure~\ref{fig:ext-op-file-format}, we provide an example expansion operation that corresponds to the operation $op_1$ of Figure~\ref{fig:ops}. 
The only addition to the Graphviz format is that the user must specify which nodes are ports and docks by enumerating them using the keywords \texttt{port} and \texttt{dock}, respectively. 
We see that node \texttt{0} is the only port of the operation, and that nodes \texttt{2} and \texttt{3} are its docks.

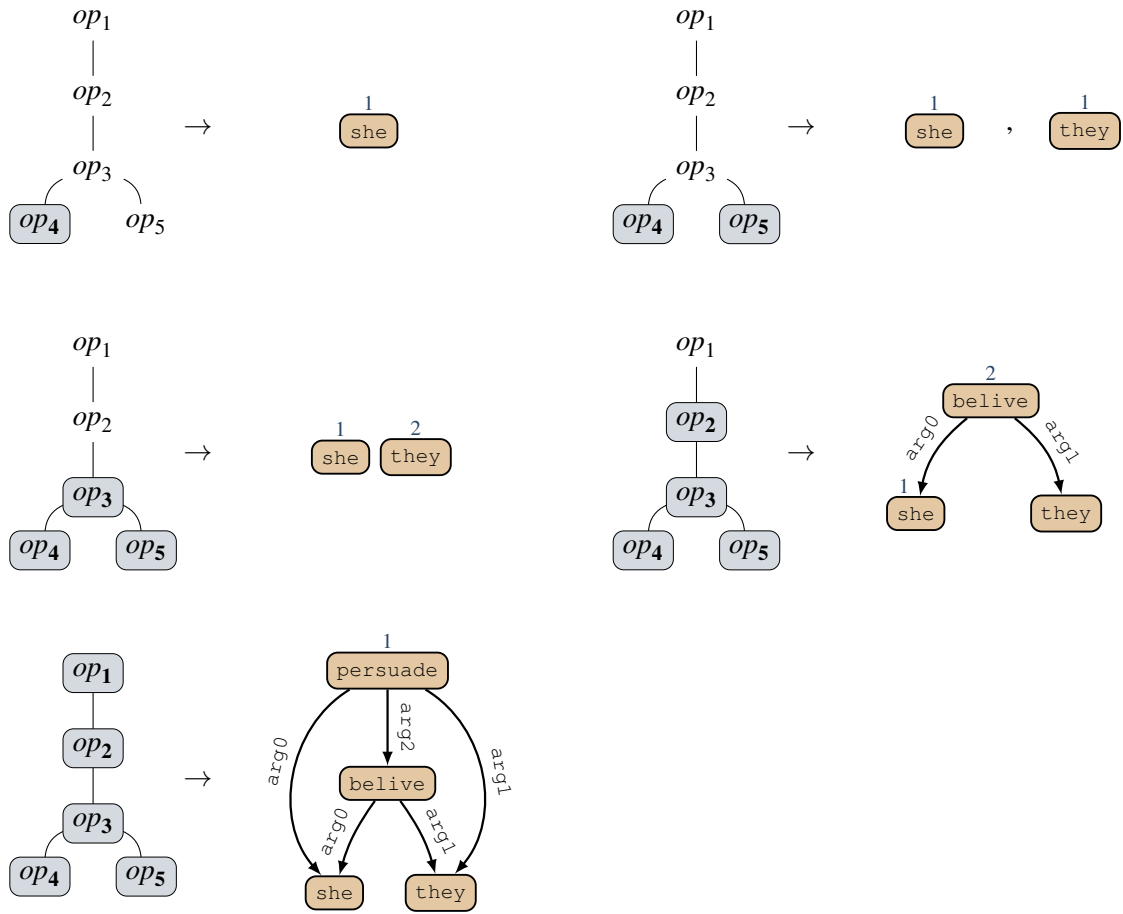
\begin{figure}[t!]
\centering
%
%
%
%
%
\tikzset{
  every label/.style={draw=none, fill=none, inner sep=1pt}
}

\newcommand{\rhs}[2]{\left.\left.\begin{array}{@{}l@{}}#1\end{array}\right[#2\right]}

\[\begin{array}{@{}l@{\ }c@{\hspace{-1ex}}c@{\hspace{1cm}}l@{\ }c@{\hspace{-1ex}}c@{}}
     \begin{tikzpicture}
    \node at (0,0) (n1) {$\mathit{op}_1$};
    \node[below of=n1] (n2) {$\mathit{op}_2$};
    \node[below of=n2] (n3) {$\mathit{op}_3$};
    \node[eval,below left of=n3] (n4) {$\mathbf{\mathit{op}_4}$};
    \node[below right of=n3] (n5) {$\mathit{op}_5$};
    \draw (n1) to (n2);
    \draw (n2) to (n3);
    \draw (n3) to[bend right] (n4);
    \draw (n3) to[bend left] (n5);
    \end{tikzpicture}
    &
    \raisebox{3.8em}{$\rightarrow$}
    &
    \begin{tikzpicture}
    \useasboundingbox (-1,-1.5) rectangle (1,2);
    \node at (0,0) [terminal, label=above:\prt 1] (y1) {\scriptsize \texttt{she}};
    \end{tikzpicture}
    &
     \begin{tikzpicture}
    \node at (0,0) (n1) {$\mathit{op}_1$};
    \node[below of=n1] (n2) {$\mathit{op}_2$};
    \node[below of=n2] (n3) {$\mathit{op}_3$};
    \node[below left of=n3,eval] (n4) {$\mathbf{\mathit{op}_4}$};
    \node[below right of=n3,eval] (n5) {$\mathbf{\mathit{op}_5}$};
    \draw (n1) to (n2);
    \draw (n2) to (n3);
    \draw (n3) to[bend right] (n4);
    \draw (n3) to[bend left] (n5);
    \end{tikzpicture}
    &
    \raisebox{3.8em}{$\rightarrow$}
    &
    \begin{tikzpicture}
    \useasboundingbox (-1,-1.5) rectangle (1,2);
    \node at (0,0) [terminal, label=above:\prt 1] (y1) {\scriptsize \texttt{she}};
    \node [right of=y1] (y2) {,};
    \node [terminal, right of=y2, label=above:\prt 1] (y3) {\scriptsize \texttt{they}};
    \end{tikzpicture}
\\\\[-8pt]
	     \begin{tikzpicture}
    \node at (0,0) (n1) {$\mathit{op}_1$};
    \node[below of=n1] (n2) {$\mathit{op}_2$};
    \node[below of=n2,eval] (n3) {$\mathbf{\mathit{op}_3}$};
    \node[below left of=n3,eval] (n4) {$\mathbf{\mathit{op}_4}$};
    \node[below right of=n3,eval] (n5) {$\mathbf{\mathit{op}_5}$};
    \draw (n1) to (n2);
    \draw (n2) to (n3);
    \draw (n3) to[bend right] (n4);
    \draw (n3) to[bend left] (n5);
    \end{tikzpicture}
    &
    \raisebox{3.8em}{$\rightarrow$}
    &
    \begin{tikzpicture}
    \useasboundingbox (-.25,-1.5) rectangle (1,2);
    \node at (0,0) [terminal, label=above:\prt 1] (y1) {\scriptsize \texttt{she}};
    \node [terminal, right of=y1, label=above:\prt 2] (y2) {\scriptsize \texttt{they}};
    \end{tikzpicture}
    &
	     \begin{tikzpicture}
    \node at (0,0) (n1) {$\mathit{op}_1$};
    \node[below of=n1,eval] (n2) {$\mathbf{\mathit{op}_2}$};
    \node[below of=n2,eval] (n3) {$\mathbf{\mathit{op}_3}$};
    \node[below left of=n3,eval] (n4) {$\mathbf{\mathit{op}_4}$};
    \node[below right of=n3,eval] (n5) {$\mathbf{\mathit{op}_5}$};
    \draw (n1) to (n2);
    \draw (n2) to (n3);
    \draw (n3) to[bend right] (n4);
    \draw (n3) to[bend left] (n5);
    \end{tikzpicture}
    &
    \raisebox{3.8em}{$\rightarrow$}
    &
    \begin{tikzpicture}
    \useasboundingbox (-1.5,-1) rectangle (2,3);
    \node at (1,1.25) [terminal,  label=above:\prt 2] (x1) {\scriptsize \texttt{belive}};
     
    \node at (0,-.25) [terminal, label=100:\prt 1] (y1) {\scriptsize \texttt{she}};
    \node at (2,-.25) [terminal] (y2) {\scriptsize \texttt{they}};
     
    \draw[->,line width=0.30mm,-latex,bend right=20] (x1) to node [pos=0.45, above, sloped] (a0) {\scriptsize \texttt{arg0}} (y1);
    \draw[->,line width=0.30mm,-latex,bend left=20] (x1) to node [pos=0.5, above, sloped] (a1) {\scriptsize \texttt{arg1}} (y2);
    \path [use as bounding box] ;
    \end{tikzpicture}%
    \\\\[-8pt]
	\begin{tikzpicture}
    \node[eval] at (0,0) (n1) {$\mathbf{\mathit{op}_1}$};
    \node[below of=n1,eval] (n2) {$\mathbf{\mathit{op}_2}$};
    \node[below of=n2,eval] (n3) {$\mathbf{\mathit{op}_3}$};
    \node[below left of=n3,eval] (n4) {$\mathbf{\mathit{op}_4}$};
    \node[below right of=n3,eval] (n5) {$\mathbf{\mathit{op}_5}$};
    \draw (n1) to (n2);
    \draw (n2) to (n3);
    \draw (n3) to[bend right] (n4);
    \draw (n3) to[bend left] (n5);
    \end{tikzpicture}  
    &     \raisebox{3.8em}{$\rightarrow$} & 
    \begin{tikzpicture} 
    \useasboundingbox (-2.5,-3) rectangle (2,1);
    \node [terminal,  label=above:\prt 1] at (0,0)  (x1) {\scriptsize \texttt{persuade}}; 
    \node [below of=x1]  [terminal, yshift=-.5cm] (x2) {\scriptsize \texttt{belive}};
    \node [below left of=x2, yshift=-.75cm] [terminal] (y1) {\scriptsize \texttt{she}};
    \node [below right of=x2, yshift=-.75cm, terminal] (y2) {\scriptsize \texttt{they}};
     
    \draw[->,line width=0.30mm,-latex,bend right=50] (x1) to node [pos=0.45, above, sloped] (a0) {\scriptsize \texttt{arg0}} (y1);
    \draw[->,line width=0.30mm,-latex,bend left=50] (x1) to node [pos=0.5, above, sloped] (a1) {\scriptsize \texttt{arg1}} (y2);
    \draw[->,line width=0.30mm,-latex] (x1) to node [pos=0.5, above, sloped] (a2) {\scriptsize \texttt{arg2}} (x2);
    
    \draw[->,line width=0.30mm,-latex,bend right=10] (x2) to node [pos=0.55, above, sloped] (aa1) {\scriptsize \texttt{arg0}} (y1);
    \draw[->,line width=0.30mm,-latex,bend left=10] (x2) to node [pos=0.55, above, sloped] (aa2) {\scriptsize \texttt{arg1}} (y2);
    \end{tikzpicture}
    
\\\\[-8pt]

\end{array}\]
\vspace{-3ex}
    \caption{
    The bottom-up evaluation of the tree in Figure~\ref{fig:tree} produced by the regular tree grammar in Figure~\ref{fig:rtg} into a graph, using the operations defined in Figure~\ref{fig:ops}. When the operation corresponding to a node in the tree is applied, the node is marked to make the derivation process clearer.}
    \label{fig:derivation}
\end{figure}

Once we have both a file specifying the operations and a file of derivation trees, we can input them to \lovelace{} by using the mandatory parameters \texttt{-g} and \texttt{-t}, respectively. 
An example usage of \lovelace{} is thus given by
\begin{center}
\texttt{java lovelace.java -g file-of-operations.txt -t file-of-trees.txt}
\end{center}
\lovelace will then evaluate the trees into graphs (by interpreting the nodes of the trees as graph operations) and output them.
The process of evaluating the tree in Figure~\ref{fig:tree} with respect to the operations in Figure~\ref{fig:ops} is depicted in Figure~\ref{fig:derivation}.
Each output graph is saved as a single text file in \texttt{gv} format (one such file resulting from our running example is depicted in Figure~\ref{fig:text_of_dag}), which makes their visualisation by Graphviz easy.
We recommend using the Graphviz online tool\footnote{\url{https://dreampuf.github.io/GraphvizOnline/}} for quick and easy graph visualisation. 
An example of a visualisation of the graph in Figure~\ref{fig:text_of_dag} by Graphviz is shown in Figure~\ref{fig:DAG_viz}.

In addition to its basic functionality, \lovelace{} allows the user to generate graphs with abstract labels, which are then replaced by concrete labels when the generated graphs are outputted. More precisely, the user can provide definitions of one-to-many label replacements, and the system will then output all possible instantiations based on these replacements. 
Such definitions are provided in a text file passed as an argument to \lovelace via the \texttt{-d} option.
This text file should include definitions for every label that shall be replaced by one or more labels. 
For example, we can generate graphs with the abstract label \texttt{sing-pronoun} which is then replaced by singular pronouns to generate various valid semantic graphs from a single result of the generation process.
As a more sophisticated example, we can expand an abstract label representing the VerbNet class \texttt{conjecture-29.5-1} (which contains, for example, the verb \texttt{believe}) by any verbs in the same class. 
In Figure~\ref{fig:definitions}, we provide a definition file in which the concepts \texttt{they}, \texttt{she} and \texttt{believe} have been expanded to provide a richer variety of semantic graphs.
When such a file is provided to \lovelace{}, all combinations of the replacements are used to create more semantic graphs. 
This naturally yields a combinatorial explosion, which is why this option should be used with care.
An alternative way of instantiating graphs is discussed in Section~\ref{sec:conclusion}.

\begin{figure}[ht]
    \centering\begin{myframe}[width=.9\textwidth,top=5pt,bottom=5pt,left=10pt,right=10pt]
\parbox[c]{\textwidth}{
\begin{lstlisting}[xleftmargin=2em,numbers=left,stepnumber=1,breaklines=true]
conjecture-29.5-1 = presume trust guess believe 
sing-pronoun = they he she
\end{lstlisting}
    }
    \end{myframe}
    \caption{Example definition file.}
    \label{fig:definitions}
\end{figure}

The three remaining parameters of the program are quite straight-forward: \texttt{-L} specifies the minimum number of nodes that a generated graph can have, \texttt{-H} is similar but instead provides an upper bound of nodes, and \texttt{-k} takes an operation name as an argument and forces every generation to use that particular operation at least once.
In Section~\ref{sec:conclusion}, we discuss other potential parameters for fine-tuning the output data that a user might be interested in.

\begin{figure}
    \centering
\begin{myframe}[width=.9\textwidth,top=5pt,bottom=5pt,left=10pt,right=10pt]

\parbox[c]{\textwidth}{
\begin{description}[font=\normalfont\ttfamily]
  \setlength\itemsep{.5ex}
    \item[\texttt{-t <tree file>}] Mandatory parameter whose argument should be the file that lists the derivation trees.
    \item[\texttt{-g <operation file>}] Mandatory parameter whose argument specifies the input expansion operations. Note that for every label of the input regular tree grammar file, there should be exactly one expansion operation specified.
    \item[\texttt{-L <min num nodes>}] Sets the minimum number of nodes in the tree. For instance, if the argument is 4, then the program skips every derivation tree that has less than 4 nodes.
    \item[\texttt{-H <max num nodes>}] Sets the maximum number of nodes in the tree. Its functionality is analogous to that of \texttt{L}'s, but it sets an upper bound of nodes instead of a lower bound.
    \item[\texttt{-d <definition file>}] This argument allows the user to replace labels in the generated graphs to create the combination of all defined label replacements.
    \item[\texttt{-k <operation name>}] Only generates the graphs whose generation process includes the specified operation. An example usage is \texttt{-k op3}.
\end{description}
}
\end{myframe}
\caption{Parameter cheat sheet.}
\label{list:parameters}
\end{figure}

To summarise the above information, we have collected the parameters implemented thus far in a cheat sheet, see Figure~\ref{list:parameters}.

\section{Conclusion and Future Work} \label{sec:conclusion}

We have presented the software \lovelace{} that generates corpora of semantic graphs; it is based on the formalism of graph expansion grammar.
To improve the software, we would appreciate input as to what features would be useful to the natural language processing community.
Below, we list some of the currently planned improvements.

As described in the previous section, \textsc{Betty} and \lovelace are currently not integrated. The user first applies the $N$-best extraction software \textsc{Betty} to a weighted regular tree grammar and then inputs the resulting list of trees to \lovelace, together with a file specifying graph operations and other parameters, to output a graph corpus.
To make the process smoother, we plan on integrating \textsc{Betty} into \lovelace{} so that the transition between both steps happens automatically.
This integration would require \lovelace{} to take additional input parameters such as the desired size of the corpus.

The graph expansion operations used in this paper are a modified version of those used in~\cite{bjorklund-etal-2023}. In some respects they are more general, while in others they are more restricted. The major differences are the following ones:
\begin{itemize}
\item Graph expansion operations are allowed to contain repetitions in the sequence of docks. This ensures that graph expansion grammars can generate all hyperedge replacement languages. For a generation system such as \lovelace, this is desirable whereas in the context of \cite{bjorklund-etal-2023}, which focuses on parsing, it is detrimental as it implies that NP-complete graph languages can be generated.
\item In this paper, context nodes can be mapped to arbitrary nodes in the argument graph of a graph expansion operation, provided that labels match. In~\cite{bjorklund-etal-2023}, admissible mappings are specified by counting monadic second-order logic, which is a much more powerful mechanism not yet implemented in \lovelace.
\item Finally, we do not make use of the mechanism of \emph{cloning} context nodes. The reason is that we consider expansion operations rather than the more restricted graph extension operations (cf.~the earlier discussion of requirements~\ref{sources} and~\ref{forget}). The former can implement cloning by iterated application of rules, which makes the use of this concept unnecessary. However, cloning may be added as an optional feature in the future to enable the user to make the rule set more compact. 
\end{itemize}

Another planned area of improvement concerns the implementation of mapping the context nodes to nodes in the argument graph. Currently, this is done by randomly choosing a node in the argument graph with a matching label.
If there is no matching candidate, then the expansion operation cannot be applied, and the program returns an error message. This is a deviation from the formal definition in two ways. On the one hand, only a single graph is returned, even though there may in fact be several results due to the nondeterminism in the formal definition. Second, if there is no matching candidate at all, the tree should simply contribute zero resulting graphs to the generated corpus instead of producing an error message. (A warning message should, however, be issued as the situation may indicate a modelling error.) One possibility would be to implement an option to have all of the formally generated graphs being outputted. Of course, this may in general cause a combinatorial explosion, similarly to how instantiating nodes using a definition file may result in a combinatorial explosion.

The possibility, mentioned above, to use logical formulas to guide the mapping of context nodes is not the only way in which to improve the user control of the context node mapping. In fact, such a control mechanism may be seen as an independent module which can be implemented in whatever way suitable. In particular, we are planning to study ways of instantiating it by neural mechanisms, which would make it possible to use machine learning to learn valid context node mappings.

Finally, future work includes developing more options to more easily fine-tune the graph generation.
There are plenty of ways in which we could extend the parameters that can be used to tweak the semantic graph corpora. 
One idea is to fine-tune the \texttt{-d} parameter: one may want the system to pick a random concept from each definition set for each instance of the concept instead of using all of the combinations.
Another idea is to control more in detail what concepts should show up in the output corpus and also to what extent, that is, a type of filtering of the corpus. 
There are several options to achieve such filtering functionality, and future work will investigate these possibilities.

\subsection*{Acknowledgements}\label{sec:Acknowledgments}
We are grateful to the anonymous reviewers for their insightful and constructive comments which helped improve the quality of this article.

\bibliographystyle{eptcs}
\bibliography{references}

\end{document}